# Testing an Einstein's intuitive objection to quantum mechanics

**Sergey A. Emelyanov**

*Ioffe Institute, 194021 St. Petersburg, Russia*

E-mail: sergey.emelyanov@mail.ioffe.ru

**Abstract.** We propose an experiment that allows one to test the Einstein's intuitive objection to Bohr's quantum mechanics (QM), which was that if QM is correct, then there should be a nonlocality related to the collapse of a *single-particle* macroscopic wavefunction, which by no means is compatible with special relativity. The idea of the experiment is related to the so-called integer quantum Hall (IQH) systems known to have *macroscopic* quantum orbits often called extended states. The experiment appeared realizable in a modified IQH system and we have found that a single-particle nonlocality does exist precisely as it follows from QM. This fact makes one come back to the Bell-Popper idea to revive the Lorentz-Poincare "dynamic" version of relativity together with the classical concept of space and time. But now, if to add the single-particle nonlocality to the Bohm-Hiley model of undivided universe, one can solve the most formidable problem – to find a preferred reference frame without involving the vague notion "aether". Moreover, now one can adopt a deeper-than-relativistic concept of what we call "reality", which opens the door to a realistic interpretation of QM and, after all, to the awareness of the worldview of this extremely successful theory.

**Keywords:** Quantum nonlocality; Quantum realism

## Introduction

When we are talking about the Einstein's physical arguments against Bohr quantum mechanics (QM), we usually mean the thought experiment with distant entangled particles, which was proposed in the well-known article by Einstein, Podolsky and Rosen (EPR) of 1935. [1]. Today the EPR nonlocality is generally regarded as a quintessence of Einstein's physical arguments against QM.

But, strictly speaking, this is not quite so. To make sure of this, let us take a closer look at the Einstein's speech at the general discussion of the Solvay Conference of 1927. There he questions the basic concept of Bohr's QM, according to which the wavefunction is an exhaustive characteristic of a quantum particle, whereas, say, in the de Broglie's version of quantum theory, a particle is additionally characterized by a spatial coordinate. Einstein argued that, if the Bohr's theory is correct, then there should be a nonlocality related to the collapse of a *single-particle* macroscopic wavefunction and this nonlocality by no means is compatible with the special relativity (SR).

To clarify his viewpoint, Einstein considers the diffraction of a single electron in a narrow slit. Since here the wavefunction propagates in a semicircle, the particle can be detected on a hemispherical screen *P* covered with a photographic film. He says literally the following:

"We can now characterise the two points of view as follows.
*1. Conception I.* The de Broglie-Schrödinger waves do not correspond to a single electron, but to a cloud of electrons extended in space. The theory gives no information about individual processes, but only about the ensemble of an infinity of elementary processes.
*2. Conception II.* The theory claims to be a complete theory of individual processes. Each particle directed towards the screen, as far as can be determined by its position and speed, is described by a packet of de Broglie-Schrödinger waves of short wavelength and small angular width. This wave packet is diffracted and, after diffraction, partly reaches the film *P* in a state of resolution.

According to the first, purely statistical, point of view $|\psi|^2$ expresses the probability that there exists at the point considered *a particular* particle of the cloud, for example at a given point on the screen.

According to the second, $|\psi|^2$ expresses the probability that at a given instant *the same* particle is present at a given point (for example on the screen). Here, the theory refers to an individual process and claims to describe everything that is governed by laws.

The second conception goes further than the first, in the sense that all the information resulting from I results also from the theory by virtue of II, but the converse is not true…

But … I have objections to make to conception II. The scattered wave directed towards *P* does not show any privileged direction. If $|\psi|^2$ were simply regarded as the probability that at a certain point a given particle is found at a given time, it could happen that *the same* elementary process produces an action *in two or several* places on the screen. But the interpretation, according to which $|\psi|^2$ expresses the probability that *this* particle is found at a given point, assumes an entirely peculiar mechanism of action at a distance, which prevents the wave continuously distributed in space from producing an action in *two* places on the screen.

In my opinion, one can remove this objection only in the following way, that one does not describe the process solely by the Schrödinger wave, but that at the same time one localises the particle during the propagation. I think that Mr de Broglie is right to search in this direction. If one works solely with the Schrödinger waves, interpretation II of $|\psi|^2$ implies to my mind a contradiction with the postulate of relativity." [2, p. 441].

Thus, Einstein considers a single-electron situation in terms of three different interpretations of quantum theory, which actually remain more or less relevant so far. The first one is the statistical interpretation (Conception I), where the wavefunction is an exhaustive description of an ensemble of particles but not a single particle. Next is the de Broglie's interpretation (Interpretation I of Conception II) that claims to describe the behavior of a single particle, but where, along with the wavefunction, the particle is also characterized by a spatial coordinate. Finally, it is the Bohr's interpretation (Interpretation II of Conception II) that claims to describe the behavior of a single particle, but where the wavefunction is an exhaustive characteristic of this particle.

Further, relying mainly on his own intuition, Einstein argues that in such a single-particle situation, it is precisely the Bohr's interpretation which should lead to a nonlocality that directly contradicts SR. This is precisely because, as it follows from Bohr's theory, the same electron, being smeared over a macroscopic volume in accordance with its wavefunction, instantaneously collapses into a small point on the screen under the detection.

As a response to the Einstein's speech, Bohr says literally the following:

"I feel myself in a very difficult position because I don't understand what precisely is the point which Einstein wants to [make]. No doubt it is my fault…
As regards general problem I feel its difficulties. I would put problem in other way. I do not know what quantum mechanics is. I think we are dealing with some mathematical methods which are adequate for description of our experiments. Using a rigorous wave theory, we are claiming something which the theory cannot possibly give … we [have] interaction [between object and measuring instrument] and thereby we put us on a quite different standpoint than we thought we could take in classical theories…The saying that spacetime is an abstraction might seem a philosophical triviality but nature reminds us that we are dealing with something of practical interest…I think the whole thing lies [therein that the] theory is nothing else [but] a tool for meeting our requirements…" [2, p.442].

So, here Bohr repeats the basic principles of the Copenhagen interpretation of quantum theory, according to which this theory is rather a set of mathematical methods to predict the results of experiments.

Any attempt to see something more behind quantum formalism is, to his opinion, misleading. He also makes it clear that if Einstein would like to show an incompleteness of QM through an opposition of QM and SR, then his intuitive guess should, as a minimum, be embodied in a particular thought experiment that clearly demonstrates where is a contradiction.

Perhaps taking into account the Bohr's position, a few years later Einstein did present a particular thought experiment. It is precisely the well-known EPR experiment. But, strictly speaking, the EPR nonlocality differs from that discussed at the Solvay Conference (this fact noted also in [3]). The point is in the EPR situation we are dealing with a two-particle wavefunction but not with a single-particle one and, in fact, this difference is of a fundamental character as the collapse a two-particle wavefunction does not necessary implies a nonlocal signaling whereas the collapse of a single-particle wavefunction is, in itself, a nonlocal signaling. As a result, at least as it is generally believed, Bohr managed to parry the EPR argument on the basis of the Copenhagen approach and thereby to avoid a direct collision of QM and SR. But this clearly would be absolutely impossible in the case of a nonlocality caused by the collapse of a single-particle wavefunction.

Of course, Einstein himself was not satisfied with Bohr's response, since he viewed the Copenhagen interpretation as a rather clever trick to avoid the problems. For example, in a letter to Erwin Schrödinger, he wrote the following: *"...The Heisenberg-Bohr tranquilizing philosophy – or religion? – is so delicately contrived that, for the time being, it provides a gentle pillow for the true believer from which he cannot very easily be aroused..."* [4]. Nevertheless, as the Bohr's (or standard) QM demonstrated more and more successes in predicting the results of a wide variety of experiments, the scientific community was more and more confirmed in the idea that the purely utilitarian approach of the Copenhagen interpretation of QM is actually the only possible one.

As for the further history of the problem, in the view of the general trend towards the adoption of Copenhagen interpretation, it was almost unnoticed that the de Broglie's realistic version of quantum theory was rediscovered by David Bohm in the early 50s [5-6]. Perhaps such an unenthusiastic reaction to this event was primarily due to the Einstein's position. In a letters to Max Born, he calls the Bohm's theory literally "too cheap" because of a deterministic description of quantum effects [7]. Nevertheless, in all experimentally-accessible (at that time) situations, including the EPR one, this theory gives exactly the same predictions as the standard QM. Therefore, the de Broglie-Bohm (dBB) pilot-wave theory is currently in the status of an alternative interpretation of quantum theory, which no one has refuted yet (for a review, see, e.g., [8]).

Today most physicists believe that the problem of how to interpret quantum theory was finally solved in the early 80s, when EPR nonlocality was demonstrated through statistical measurements in the EPR situation [9]. Such measurements are known as Bell test as they are based on the so-called Bell theorem of 1964 [10]. This test is believed to finally refute the dBB theory together with the other potential hidden-variable theories because de facto these theories recognize a real (though non-causal) correlation between the past and the future. For the same reason of actual recognition of such a correlation, almost no one notes currently the quantum model of undivided Universe, which was proposed by David Bohm together with Basil Hiley in 1993 [11].

But a quite unexpected (for most physicists) conclusion from the Bell test was made independently by John Bell himself as well as by the philosopher Karl Popper [12-13]. They believed that that the sacrificing of physical realism is too much a price we should pay to preserve the relativistic view of space and time. For this reason, they both proposed to abandon the Einstein's "kinematic" version of relativity and to return to the Lorentz-Poincare "dynamic" version with its classical concept of space and time.

At that time, however, this idea was not taken seriously for at least two reasons. Firstly, in the minds of most physicists, the Lorentz-Poincare theory inevitably leads to the problem of the so-called "aether" as an immobile medium that should play the role of a preferred reference frame. It is precisely the problem that makes the Lorentz-Poincare theory much less elegant than SR and therefore the return to this problem looks like a step back for the entire physics. Secondly, at that time, the only realistic version of quantum theory was the dBB theory. As a result, most physicists believed that to preserve quantum realism we inevitably should reject QM in favor of the dBB theory. But such a revolutionary step seems, as a minimum, unjustified at least because the predictions of QM are always consistent with experiments. And although the dBB theory also never contradicted experiments, it is generally regarded as a "too classical" perhaps because of its deterministic character.

Thus, although the Bell test allows for various interpretations, it currently is regarded as a strong argument in favor of a purely utilitarian approach to QM and accordingly against the Einstein's idea concerning the incompatibility of QM and SR together with his philosophical concept of the so-called objective reality as the ultimate subject of science.

**Main part**

However, going back to the Bohr-Einstein times, the question arises why, after the 1927 Solvay Conference, as an argument against QM, Einstein used the EPR situation with the collapse of a two-particle wavefunction but never mentioned the situation with the collapse of a single-particle wavefunction even though the latter would make meaningless any Bohr's attempts to deny a direct conflict between QM and SR. It seems the point is the situation that was discussed at the Conference, is, strictly speaking, not an argument but rather an intuitive objection. To be a real argument, this objection needs to be embodied, as a minimum, in a more or less plausible thought experiment. But actually not so many physical systems were known at that time where quantum effects would manifest themselves on a macroscopic scale and moreover at the level of individual particles. It may well be that it is extremely hard to find among them a system where the collapse of a single-particle wavefunction would lead to an observable nonlocality.

But now, in the context of this task, it seems interesting to consider the macroscopic quantum effects revealed not so long ago. To do that, it should be noted that perhaps the most successful in revealing of such effects is the solid-state physics. We mean the well-known superconductivity together with related effects, as well as a less known but no less interesting effect called the quantum Hall effect.

At first glance, solid-state systems are hardly relevant in this context since, as a rule, they have a huge number of electrons and therefore one would expect rather collective quantum effects than the effects at the level of individual particle. This is more or less correct for the superconductivity and for the so-called fractional quantum Hall effect. However, if one takes what is known as the integer quantum Hall (IQH) effect, here we unexpectedly are dealing with spatially separated and hence distinguishable electrons, whose behavior should be regarded not at a statistical level but at the level of individual particles.

Leaving aside (for now) the details, here we only note that the IQH effect is a low-temperature macroscopic quantum effect that manifests itself in an exact quantization of transverse conductivity of two-dimensional electronic systems in a strong magnetic field. The discovery of this effect by Klaus von Klitzing was awarded the Nobel Prize in 1985 [14]. Ultimately, the effect is related to the spatially-separated quantum states near the system edges, in which the electron wavefunction is reminiscent a quasi-one-dimensional orbit stretched along the perimeter of the entire macroscopic system (for a review see, e.g., [15]). These states were first described theoretically by Bertrand Halperin in 1982 and he called them extended (or current-carrying) states [16].

Certainly, such object as a macroscopic orbit-like quantum state was unknown at the Bohr-Einstein times. Therefore, it seems interesting to take the first step, that is, by using of this object to give the Einstein's objection the form of a thought experiment where the collapse of a single-particle wavefunction would lead to an observable nonlocality.

***Thought experiment with a nonlocality caused by the collapse of a single-electron wavefunction***

To this aim, let us consider the following situation. Suppose we have a flat quasi-one-dimensional macroscopic orbit (*O*). For certainty, let it be a square with a side *d* because in IQH systems the shape of the orbit is determined only by the shape of the entire system (Fig. 1). As it is characteristic for any IQH system, we suppose that all space, in which the orbit is, is filled by the local charged scatterers, the density of which is so high that the electron's mean free path is much less than the lengthscale of the orbit and therefore here electron in no way can overcome any macroscopic distance by means of a continuous motion. Accordingly, the electron's lifetime in the orbit ($\tau$) is determined by the interaction with these local scatterers that transfer it into a vacant state near the orbit. Finally, suppose that near the left side of the orbit there is a deep level *A*, from which electron can be transferred into the orbit *O* by a strictly local excitation by the photons of a proper energy.

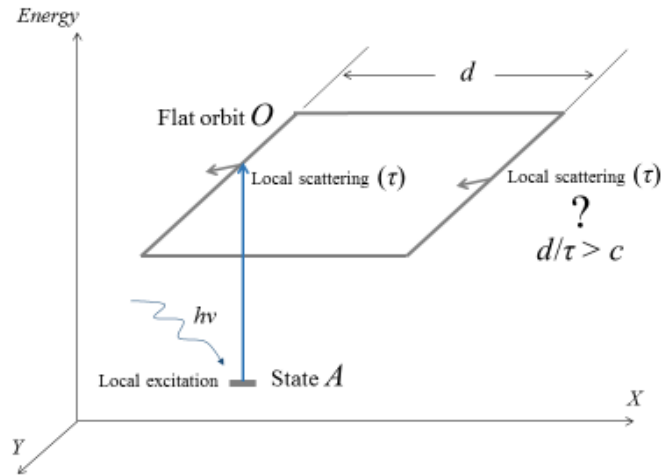

**Fig. 1**. Thought experiment with the collapse of a single-electron macroscopic wavefunction in an IQH system.

Consider now what is the further behavior of the electron according to the dBB theory (Interpretation I of the Conception II) as well as according to the standard QM (Interpretation II of the Conception II). We will not consider the Conception I simply because it prohibits any analysis at the level of individual particles. Then we have the following two versions:

Version No. 1 (dBB theory). Along with the wavefunction, electron always has a spatial coordinate and therefore it always moves along some trajectory.

Here the behavior of the electron is quite trivial. Since it cannot move along any macroscopic trajectory due to the high density of the scatterers, it can only be detected very close to the spatial region where it was excited. Therefore, as Einstein suspected, here there is no any nonlocality.

Version No. 2 (standard QM). The wavefunction is an exhaustive description of the electron.

Here we face a completely different picture that seems even a bit paradoxical. Since the electron does not have any definable coordinate in the orbit, the probability of the scattering (or equivalently detection) near the right side of the orbit is exactly the same as the probability of the detection near the left side regardless of the lengthscale of the orbit. This means that at least with a probability of one quarter, the electron will overcome the distance *d* during the time $\tau$. But insofar as the distance *d* is actually unrestricted, we always can provide $d/\tau > c$, where *c* is the speed of light. Moreover, the probability of this nonlocal effect can easily be enhanced, for example, by a higher concentration of scatterers near the right side.

An important characteristic feature of the situation is that, to prove nonlocality related to the collapse of single-electron wavefunction, we need only to find out electron near the right side of the orbit. The high concentration of scatterers guarantees us that here electron may emerge only by a quantum jump at a distance of no less than d, because any continuous movement over a macroscopic distance is impossible, at least if we neglect an extremely slow diffusion which moreover has no a preferred direction.

Thus, Einstein was quite right when he intuitively suspected that if wavefunction is an exhaustive characteristic of a particle, then there should be a nonlocality due to the collapse of a single-particle wavefunction, which implies a nonlocal signaling. Therefore, if our thought experiment is realizable, then there is a dilemma either Bohr's QM or Einstein's SR. *Tertium non datur*.

## IQH system as a potential tool to realize the thought experiment

Of course, the above dilemma is of a revolutionary character and therefore it seems extremely interesting to realize the thought experiment. To assess the possibility of this in the IQH system, consider this system in more detail.

In fact, typical IQH system is a two-dimensional semiconductor structure where the thickness of conducting layer is less than the electrons' mean free path (a few tens of nanometers). Therefore, in two dimensions (*X* and *Y*), electrons are free with the wavefunction of a Bloch type while in the third dimension (*Z*) their energy spectrum is a series of quantum levels. To reach the IQH regime, at a low temperature, one applies a strong magnetic field in the Z-direction so that the electron motion becomes quantized in all directions, i.e. each quantum-size level gives rise to a series of sublevels known as Landau levels.

Fig. 1 (left panel) shows the energy diagram of an idealized IQH system where we neglect the local potential fluctuations. Here the *X*-direction is arbitrary because of the axial symmetry in the *XY* plane. Most electrons are localized within the so-called cyclotron orbits and their absolute number in each Landau level is determined by the dense packing of these orbits in the *XY* plane. The electrons are strongly degenerate with respect to their wave vector in the *Y* direction ($k_y$) or equivalently with respect to the *X*-coordinate of the center of their cyclotron orbits ($X_o$) because of the following relation: $X_o = -\lambda^2 k_y$, where $\lambda$ is the so-called magnetic length.

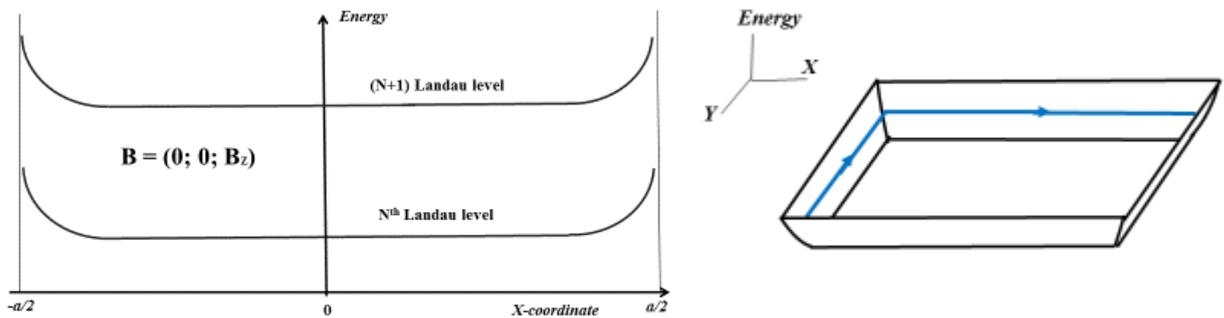

**Fig. 1**. Left panel: Energy diagram of an idealized IQH system where the local potential fluctuations are neglected. The scale is strongly exaggerated near the edges to show a strip with orbit-like quantum states stretched along the perimeter of the system. Right panel: Energy diagram of a single Landau level in two dimensions. One of the macro-orbits is shown in blue.

However, within a microscopic strip near the edges there is always a strong in-plane electric field perpendicular to the edge. As a result, here electrons are in a crossed electric and magnetic field that lifts the Landau level degeneracy and gives rise to spatially-separated orbit-like quantum states stretched along the perimeter. These are precisely the Halperin's extended states with a characteristic cross-section of the order of cyclotron radius. Their spatial configuration is shown in the right panel of Fig. 1. The system thus looks like a bowl with a large flat bottom and narrow walls filled by macro-orbits.

At first glance, this system is suitable to realize the test, if one provides a strictly local photoexcitation of electrons between the orbits belonged to different Landau levels. But actually there is a serious problem. Since here the macro-orbits can exist only near the system edges, their absolute number is extremely small. As a result, the detection of any effects related to their photoexcitation is almost impossible.

A solution to this problem was proposed in [17] and it is based on the fact that most IQH system is asymmetric in the *Z* direction. We mean an asymmetry of their confining potential, which may be related, for example, to a penetration of so-called surface potential into the conducting layer. Such an asymmetry is equivalent to an electric field in the *Z* direction, which actually may reach a gigantic value of the order of $10^5$V/cm. This field is often called built-in field ($E_{buit-in}$) and it is the reason for a number observable effects.

Now if such an asymmetric structure is in a magnetic field which, along with the quantizing component ($B_Z$), has also a nonzero in-plane component ($B_X$), then a strong crossed electric and magnetic field occurs not only near the edges but also in the system interior. Situations of that kind were calculated for some limiting cases back in the early 90s [18-19]. Based on these calculations, one would suppose that here the Landau level degeneracy may be lifted throughout the whole system in the X-direction. Then the energy diagram looks like that shown Fig. 3 (left panel) whereas the energy diagram of a single level (or band) in two dimensions looks like that shown in the right panel of Fig. 3.

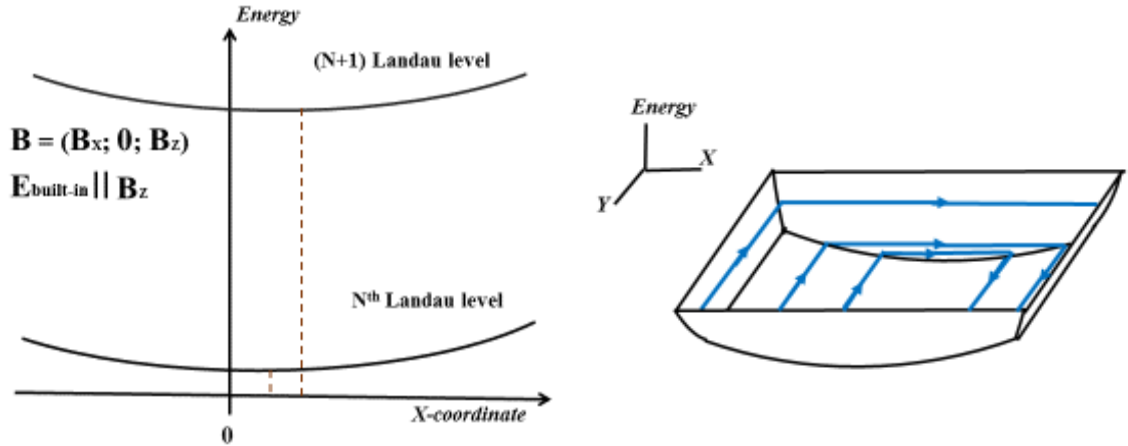

**Fig. 3**. Left panel: Energy diagram of an asymmetric IQH system in tilted quantizing magnetic field. Landau level degeneracy is lifted in the *X*-direction. The scale is strongly exaggerated near zero point to show a small shift of the bands in this direction. Right panel: Energy diagram of a Landau level in two dimensions. Orbit-like states are shown in blue.

Here the lifting of Landau level degeneracy is accompanied by a slight shift of the levels from each other in the *X*-direction as well as by the emergence of macro-orbits in the system interior as the energy diagram of each Landau level is again something like an energy bowl but now with a curved bottom along the *X* axis. Thus, the orbits fill each Landau level as it is shown in the right panel of Fig. 3 and their characteristic cross-section is of the order of the electrons' cyclotron radius determined by $B_X$.

The distinct feature of the system is the presence of an asymmetry in the *XY* plane, which can be characterized by a polar vector $B_X \times E_{built-in}$. This fact opens the door to an experimental method that may be suitable to detect macroscopic electron jumps along the *X*-axis (if any). We mean the measurements of local currents along the *Y* axis under a local excitation of this system by a terahertz laser radiation as the energy of quanta of this radiation is close to a typical energy gap between Landau levels. Phenomenologically, it is the so-called photovoltaic currents, the measurement of which has been successfully used as a method to study semiconductor systems for a long time (for a review, see [20]). But this method has never been used in the IQH systems perhaps because, as a rule, there are no current-carrying states in the interior of such systems.

### *Realization of the thought experiment in asymmetric IQH system*

Of course, before starting the test, we should demonstrate that the macro-orbits do exist in the interior of asymmetric IQH system in tilted quantizing magnetic field. But actually this has already been done in our previous work precisely by the measurements of local currents induced by terahertz laser radiation [17]. It has also been shown that the shift of the Landau bands along the *X* axis is negligibly small compared to the system size, i.e. the macro-orbits are almost symmetric with respect to the *X* = 0 (see Fig. 3).

Taking into account these results, we use the experimental scheme sketched in Fig. 4 (left panel). Here the structure is 20mm in width (*X* axis) and 16mm in length (*Y* axis). This structure is covered with a mask non-transparent for the laser radiation. The mask has four identical windows symmetrical with respect to the center of the structure (*X* = 0). Each window is 2mm in width and 12mm in length. The distance between the adjacent windows is 3mm and each active region is supplied with a pair of ohmic contacts to measure local current along the *Y* axis.

We use the structures grown by the method of molecular beam epitaxy (MBE). These are the so-called single quantum well structures of type GaSb-InAs-GaSb with the well width of 15nm. Such structures are known to have a high electron density (about $1.5 \cdot 10^{12} cm^{-2}$) as the top of the GaSb valence band is above the bottom of the InAs conduction band. To avoid the so-called electron hybridization related to this overlapping, the conducting layer is sandwiched between two thin AlSb barriers (3 nm each). In these structures, the density of point-like charged scatterers is so high that the low-temperature scattering time is as short as about 3ps and the mean free path of excited electrons is less than 0.1µm. To provide "built-in" electric field, the surface-to-well distance is about 20nm whereas the penetration depth of the surface potential is about five times more [21]. The working temperature is about 2K.

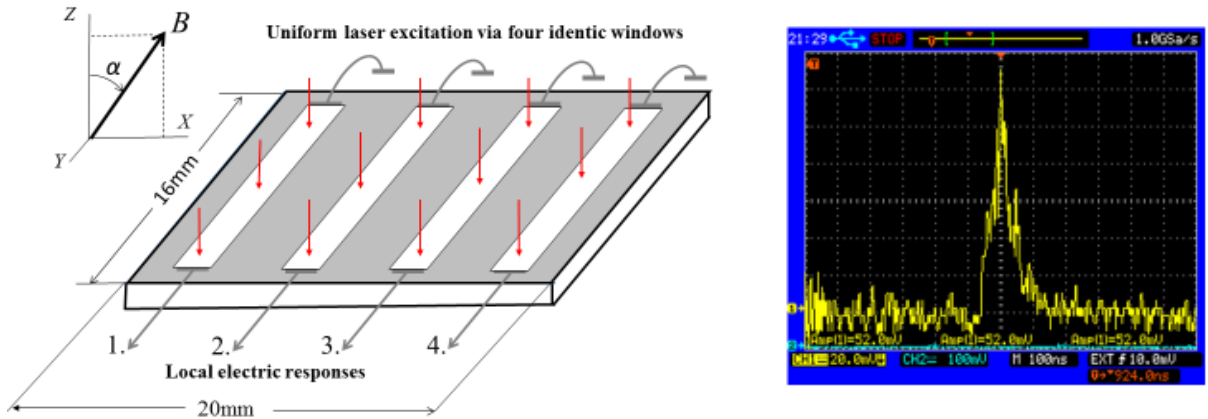

**Fig. 4.** Left panel: Sketch of the test. The structure is in tilted magnetic field and terahertz laser radiation may enter the structure through four identical windows, each of which can be either open or closed.
Right panel: Typical track of a terahertz laser pulse that excites the system (100ns/div.).

Terahertz laser pulses are provided by an optically-pumped ammonia laser. The laser wavelength is 90.6µm ($\hbar\omega$ = 13.7meV), pulse duration is about 40ns, and intensity is up to 200W/cm². Typical laser track is shown in Fig. 4 (right panel). Magnetic field of up to 6T is provided by a superconducting magnet. This field can be tilted by an arbitrary angle *α* in the *XZ* plane. We measure local currents those kinetics is similar to that of the laser pulses.

To avoid any ambiguity, we compare the system's responses in two regimes – with and without the Landau quantization. These regimes can be switched from one to another *in situ* by rotating the magnetic field in the *XZ* plane. In the no-quantization regime, magnetic field has only an in-plane component. In this case, responses may also arise insofar as the vector product $B_X \times E_{built-in}$ remains nonzero but there should be no any macro-orbits. In the regime of Landau quantization, magnetic field has both quantizing ($B_Z$) and in-plane ($B_X$) components. It is the regime with macro-orbits in the system interior. Since, as it follows from our previous experiments, the gap between the Landau bands is close to the energy of laser quanta at $B_Z \approx$ 4.8T, we use the total magnetic field of about 4.8T with a small tilting angle of about 8˚.

The procedure of the test is as follows. To find out the spatial distribution of photo-excited electrons in the *X*-direction under a strictly local laser excitation, we measure responses from *each* contact pair when *only one* of four windows is open. As the sensitivity of our method per one excited electron may be a function of the *X*-coordinate, each response will be normalized to that one observed here when all windows were open. For convenience, the responses from illuminated region will be shown in yellow while the responses from regions in the dark (if any) will be shown in blue.

The outcome of the test is presented in Fig. 5. Left panel shows the spatial distribution of local responses in the no-quantization regime. It is seen that here the picture is quite trivial. Responses arise only in the illuminated regions and they are almost the same. In other words, non-equilibrium electrons are detected only where they are excited.

But the picture changes drastically in the regime of macro-orbits (Fig. 5, right panel). The illumination of the region No.1 results in a local response not only here but also in the symmetrical region No.4 whereas there are no responses in the other regions (No.2 and No.3) even though they are much closer to the laser spot. Moreover, both responses are almost *the same* despite the fact that the distance from the laser spot to the region No.4 is *five orders higher* than the mean free path of excited electrons. Similarly, the illumination of only the region No.4 gives rise to a local response not only here but also in the region No.1 and these two responses are almost the same again.

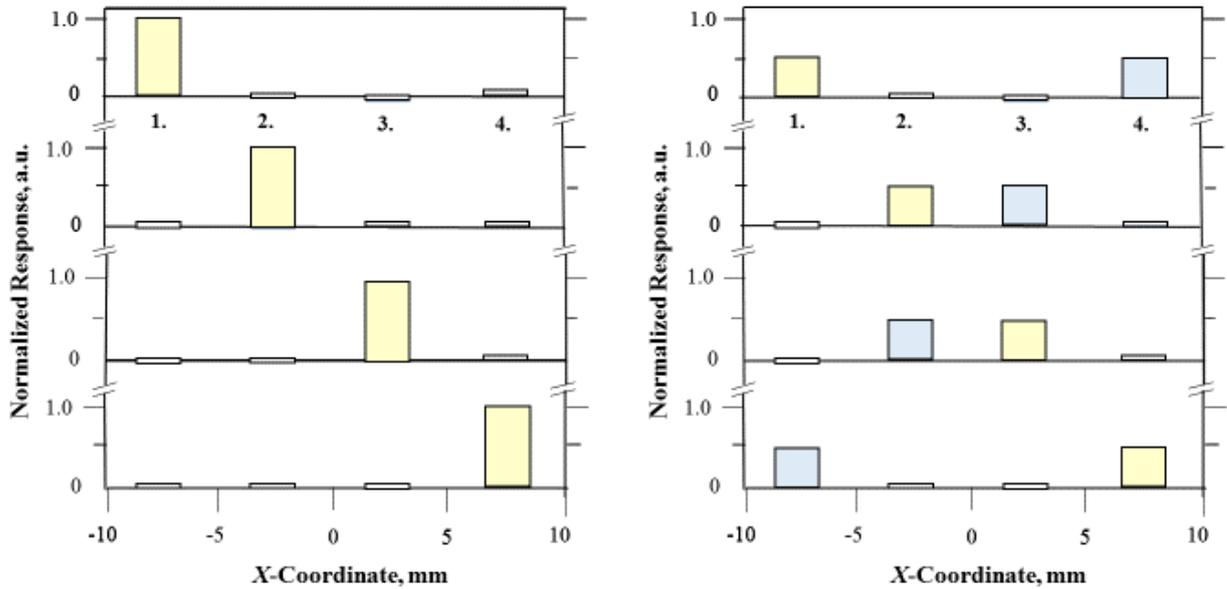

**Fig. 5.** Left panel: Spatial distribution of local responses in the no-quantization regime.
Right panel: Spatial distribution of local responses in the regime of macro-orbits.

The same picture is observed when we illuminate only the region No.2. Local response occurs not only here but also in the symmetrical region No.3 whereas there no responses in both the region No.1 and the region No.4. This fact clearly shows that the detection of excited electrons far away from the laser spot by no means can be caused by their diffusion because the region No.1 and the region No.3 are at the same distance from the laser spot. Finally, we perform a synchronous detection of responses from the region No.1 and the region No.4 when only the region No.1 is illuminated. Typical tracks are shown in Fig. 6. It is seen that there is no a delay between the responses at least within the time resolution and therefore any slow effects such as diffusion are irrelevant indeed.

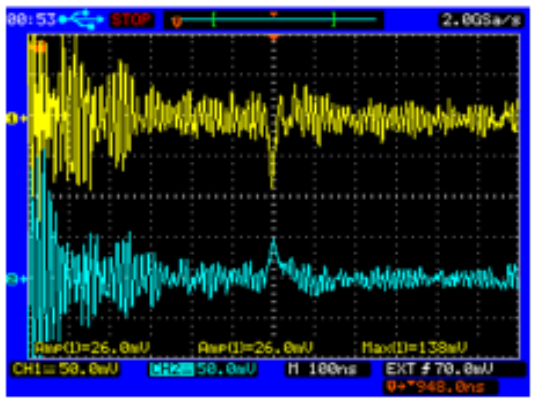

**Fig. 6.** Typical tracks under synchronous detection of local responses in the region No. 1 (upper track) and the region No. 4 (lower track) when only the window No. 1 is open in the regime of Landau quantization ($\alpha = -8°$, $B = 4.8T$). Timescale is 100ns/div.

Thus, on the one hand, our experiments confirm that Einstein was quite right suspecting that if wavefunction is an exhaustive characteristics of a quantum particle, as predicted by QM, then a nonlocal effect should exist which is caused by the collapse of a single-particle wavefunction. But, on the other hand, this argument against QM surprisingly (for him and perhaps for Bohr too) turns into an experimentally-tested argument against SR because the counterintuitive predictions of QM are fantastically coming true.

**Fundamental consequences**

*The Bell-Popper idea at a higher level of knowledge*

Actually, what we observe is nothing else but a discrete spatial dynamics (DSD) of a single particle with macroscopic discontinuity and such dynamics is truly incompatible with SR. This means that we inevitably should revive the Bell-Popper's idea to return to the Lorentz-Poincare "dynamic" version of relativity. But now the situation is more favorable for such a return than in the days of Bell-Popper because now we can do that at a higher level of knowledge, which allows one to answer the arguments against this idea, which were put forward earlier.

So, the first argument is related to the problem of preferred reference frame, which has always been a stumbling block for the Lorentz-Poincare theory. More precise, it is the problem with the vague notion "aether", which currently is regarded as an inevitable attribute of this theory. But now it becomes clear that actually it is a pseudo-problem caused by the current relativistic insight of the Universe. To see that, one should take into account the following two things. First, now we can revive the Bohm-Hiley quantum model of undivided Universe as it no more needs to be squeezed into the framework of the relativistic concept of spacetime. Secondly, DSD nonlocality is actually nothing else but the quantum concept of absolute simultaneity, which rests not on the meaningless notion of an infinite speed as in classical physics but on the experimentally-tested notion of the instantaneous collapse of macroscopic wavefunction of a single particle.

Thus, combining these two things, it is easy to see that both a true time and a true length can be established in such Universe. Accordingly, for each physical object considered separately from the entire Universe, the rest of this Universe is precisely the preferred reference frame we need. Quantum theory appears thus to be a more general theory than the theory of relativity because, in a broad sense, the former is the theory about a wholeness of the world while the latter, by definition, supposes the presence of at least two entirely independent bodies though, in terms of quantum approach, this is always a kind of approximation.

*A quantum concept of reality*

However, as for the second argument against the Bell-Popper idea, namely, that this idea rejects Bohr's QM in favor of the "too classical" dBB theory, in fact, this is also a manifestation of our relativistic insight of physical world. First of all, our experiments directly contradict the dBB theory with its continuous trajectories of quantum particles and leave only the Bohr's theory at least because it is the only theory that predicts DSD nonlocality. But the question is whether QM inevitably should be regarded as a purely utilitarian mathematical method or another route of how to interpret quantum formalism has become possible now.

To clarify this issue, we start with the fact that everyone who tried to solve the problem of interpretation of this formalism were limited by the relativistic concept of reality, which directly follows from the relativistic concept of spacetime. In the frames of this concept, reality is strictly limited by the events in three-dimensional Euclidean space or, more precisely, in four-dimensional Minkowski spacetime. Therefore, we actually have no choice other than to admit that QM describes something beyond the reality precisely because of the appeal to a multi-dimensional Hilbert space that is the space of quantum states, but not to the "real" space if reality is restricted by the Minkowski spacetime.

Actually, the problem of incompatibility of the mathematical language of QM with the relativistic insight of the world was fully realized by Einstein who, at the same 1927 Solvay Conference, pointed out that the appeal of QM to a Hilbert space is a kind of challenge to the current physical worldview. For this reason, shearing the Einstein's opinion, most participants expressed the hope that such an appeal is a kind of "growth disease" of QM, which eventually will be overcome in the sense that an improved version of this theory will appeal only to what is known as "real" space. However, as we know, this hope has not come true, because the appeal to Hilbert space has appeared a fundamental thing, without which QM merely loses its meaning.

But now, if we remove the relativistic concept of spacetime or, more precisely, if we separate time from the spatial coordinates, then the "classical" method to interpret physical theories becomes possible when behind every concept introduced by a theory we suppose a real physical substance. If we follow this method, then we should recognize that there is a real world of quantum states in which quantum particles "live" according to the quantum laws described in QM. Of course, this world is unobservable for us. But it does not mean that this world does not exist at least because it is in a permanent connection with the observable world through what we call "measurement".

Now it becomes clear why does QM imply a peculiar role of the notion "measurement" which, as stressed by Bohr, has nothing to do with what this notion means in classical physics. For by the notion "measurement", QM implies a physical process when a quantum particle becomes an integral part of a macroscopic body (for example, apparatus) and hence becomes an element of the observable world. Therefore, obeying the laws of this world, the particle no longer can be in a superposed state, as was possible in a deeper quantum reality, but should choose only one of possible states. Thus, quantum measurement is a bridge by which a particle may come to observable world, but, being separated from the macroscopic body, it again becomes an element of unobservable reality.

An important point is that, as a physical process, quantum measurement is quite objective and here the presence of an observer or any other living beings is actually not necessary. This fact was noted many years ago, for instance, by Landau and Lifshitz who, in their famous course of quantum mechanics, write literally the following: *"... we emphasize that, in speaking of 'performing a measurement', we refer to the interaction of an electron with a classical 'apparatus', which in no way presupposes the presence of an external observer"* [22]. This means that QM may well be consistent with the Einstein's philosophical concept of objective reality, without which, as he believed, science is impossible. But, to make so, the concept of reality should be broadened to include two "parallel" worlds – observable and unobservable. Certainly, each of these worlds obeys its own laws, but, on the hand, they have the same concept of time, and, under certain conditions, there may be the transitions of quantum particles from one world to another.

In fact, we thus rehabilitate the Einstein's idea of objective reality as the ultimate subject of science. But, on the other hand, this idea is now combined with the Bohr's idea that the comprehension of reality is, in a sense, the comprehension of what we could say about the reality which actually is broader than the observable world, whereas it is precisely the observation that still is the only way to investigate physical world.

Finally, it should be noted that, generally speaking, the concept of a deeper reality allows one to resolve a lot of quantum paradoxes. As an example, let us take the well-known paradox of Schrödinger cat. Here a radioactive atom, as a quantum object, belongs to a deeper reality where it may well be in a superposition of two states: excited and non-excited. At the same time, any superposed state may exist only in the deeper reality and by no means can be attributed to a macroscopic body in the observable reality. Therefore, for the cat, such a situation is always not more than a sort of the game "heads and tails" but where the probability of fatal outcome is a complicate (but certainly increasing) function of time.

### *Mystery of the worldview of QM*

Not so long ago, David Mermin uttered a phrase that seems a bit shocking by its frankness: *"If I were forced to sum up in one sentence what the Copenhagen interpretation says to me, it would be "Shut up and*

*calculate!"* [23]. But actually this is the so accurate and succinct formulation of the current insight of QM that his formula *"Shut up and calculate"* quickly spread among physicists and became a kind of meme. In fact, however, this shocking formula is fully consistent with the Bohr's idea that QM is merely a mathematical method to predict the results of experiments. Moreover, the other outstanding physicist, Lev Landau, spoke in the same spirit long before Mermin as the Landau's formula *"Think less about foundations"* may well be regarded as merely a milder expression of the Mermin's phrase [24].

However, an inner discomfort and even disappointment involuntarily arises from a purely utilitarian insight of QM because this means that the most successful theory in the history of science gives us nothing new in our worldview. This, of course, did not suit Einstein, who expressed his attitude to QM as follows: *"quantum mechanics is certainly imposing. But an inner voice tells me that it is not yet the real thing. The theory says a lot, but does not really bring us any closer to the secret of the "Old One"* [25]. Similarly, John Bell never accepted a purely utilitarian insight of QM and was sure that *"… the probing of what quantum mechanics means must continue and … it will continue, whether we agree or not that it is worthwhile…"* [12, p. 52]. Moreover, his well-known sympathy to the dBB theory as well as the sympathy to this theory of many other physicists seems primarily due to precisely the lack of any worldview in QM.

But perhaps the most profound impression is made by both the form and the content of how Richard Feynman spoke about the problem of the worldview of QM: *"We have always had a great deal of difficulty understanding the worldview that quantum mechanics represents. At least I do, because I'm an old enough man that I haven't got to the point that this stuff is obvious to me. Okay, I still get nervous with it…. You know how it always is, every new idea, it takes a generation or two until it becomes obvious that there's no real problem. I cannot define the real problem, therefore I suspect there's no real problem, but I'm not sure there's no real problem"* [26].

Mermin, for example, regards the Feynman's speech as a kind of poetry [27]. It seems this is indeed the case, because here there are no any clear statements but rather deep personal experiences characteristic for a poet rather than for a physicist. And the subject of these experiences is again the lack of any worldview in QM.

But now, in the light of DSD nonlocality which de facto frees QM from the shackles of the relativistic worldview, we are on the verge that this theory will magically transform from a purely utilitarian mathematical method into a complete fundamental theory with the deepest worldview ever proposed by physical theories. And it is precisely the status deserved by the most successful theory in history of science.

## Acknowledgements

The author is grateful to Prof. Sergey Ivanov (Ioffe Institute) for the MBE samples as well as to Prof. Raymond Chiao (UC at Merced) for useful comments on the experiment.